\documentclass[preprint,authoryear,12pt]{elsarticle}

\usepackage{epsfig, rotating, longtable, lscape, longfigure}

\usepackage{amssymb}

\usepackage[ps2pdf,%
a4paper=true,%
breaklinks=true,%
colorlinks=true,%
pdfauthor={First Author et al.},%
pdftitle={Template for manuscripts in Advances in Space Research}%
]{hyperref}

\journal{Advances in Space Research}

\begin{document}

\begin{frontmatter}



\title{\emph{Neil Gehrels Swift Observatory} studies of supersoft novae}


\author{K.L. Page}
\address{Department of Physics \& Astronomy, University of Leicester, LE1 7RH, UK}
\ead{klp5@leicester.ac.uk}


\author{A.P. Beardmore}
\address{Department of Physics \& Astronomy, University of Leicester, LE1 7RH, UK}
\ead{ab271@leicester.ac.uk}

\author{J.P. Osborne}
\address{Department of Physics \& Astronomy, University of Leicester, LE1 7RH, UK}
\ead{opj@leicester.ac.uk}

\begin{abstract}
The rapid response capabilities of the \emph{Neil Gehrels Swift
  Observatory}, together with the daily planning of its observing
schedule, make it an ideal mission for following novae in the X-ray
and UV bands, particularly during their early phases of rapid
evolution and throughout the supersoft source interval. Many novae, both classical and recurrent, have been extensively monitored by
\emph{Swift} throughout their supersoft phase and later decline. We
collect here results from observations of novae with outbursts which occurred between the start of 2006 and the end of 2017.

\end{abstract}

\begin{keyword}
  novae, cataclysmic variables; ultraviolet: stars; X-rays: stars

\end{keyword}

\end{frontmatter}

\parindent=0.5 cm

\section{Introduction}
\label{intro}

A nova occurs in a binary system, consisting of a white dwarf (WD) and a late-type main sequence or giant secondary star, when sufficient hydrogen is transferred from the secondary to the WD surface such that the temperature and pressure are high enough to ignite the material, leading to a thermonuclear runaway. After the initial explosion (which will appear as a new optical source), the ejected envelope spreads out, becoming optically thin; at this stage, the surface nuclear burning becomes visible (assuming it is still ongoing). This emission peaks in the soft X-ray band, and is therefore known as the Supersoft Source (SSS) phase.

Besides this soft emission, novae also often show faint, hard ($\sim$~1--10~keV) X-rays \citep[e.g.,][]{brecher77, orio01, muk08, chom14b}, caused by ejecta shocks; this harder emission may be detectable before, during and after the SSS phase. More recently, novae have also been detected in the GeV $\gamma$-ray range by the \emph{Fermi}-Large Area Telescope \citep[LAT;][see Section~\ref{open} for more discussion]{atwood09}.



\section{Setting the scene}

\begin{figure}
\begin{center}
  \includegraphics*[width=6.5cm]{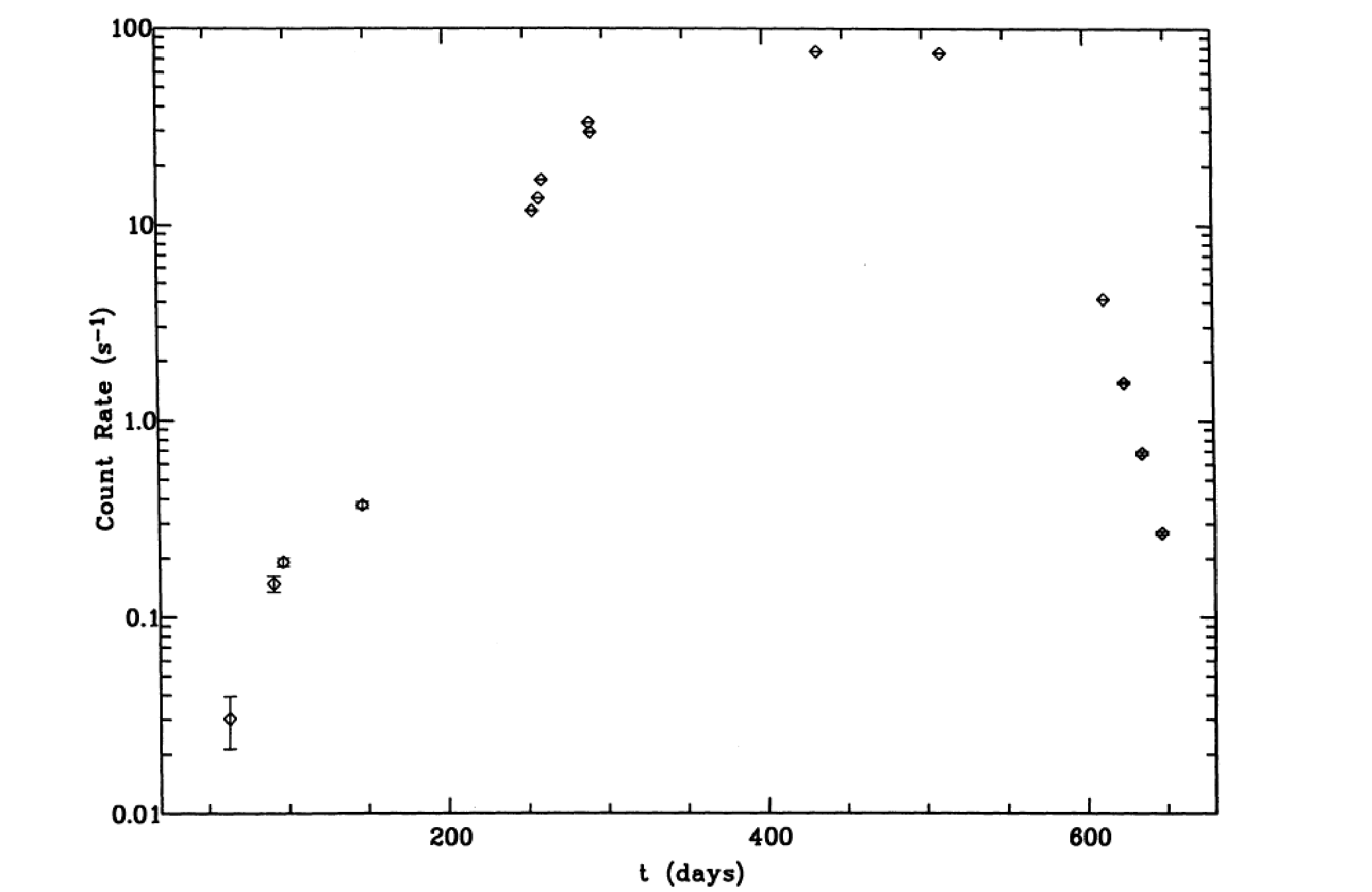}
  \includegraphics*[width=6.5cm]{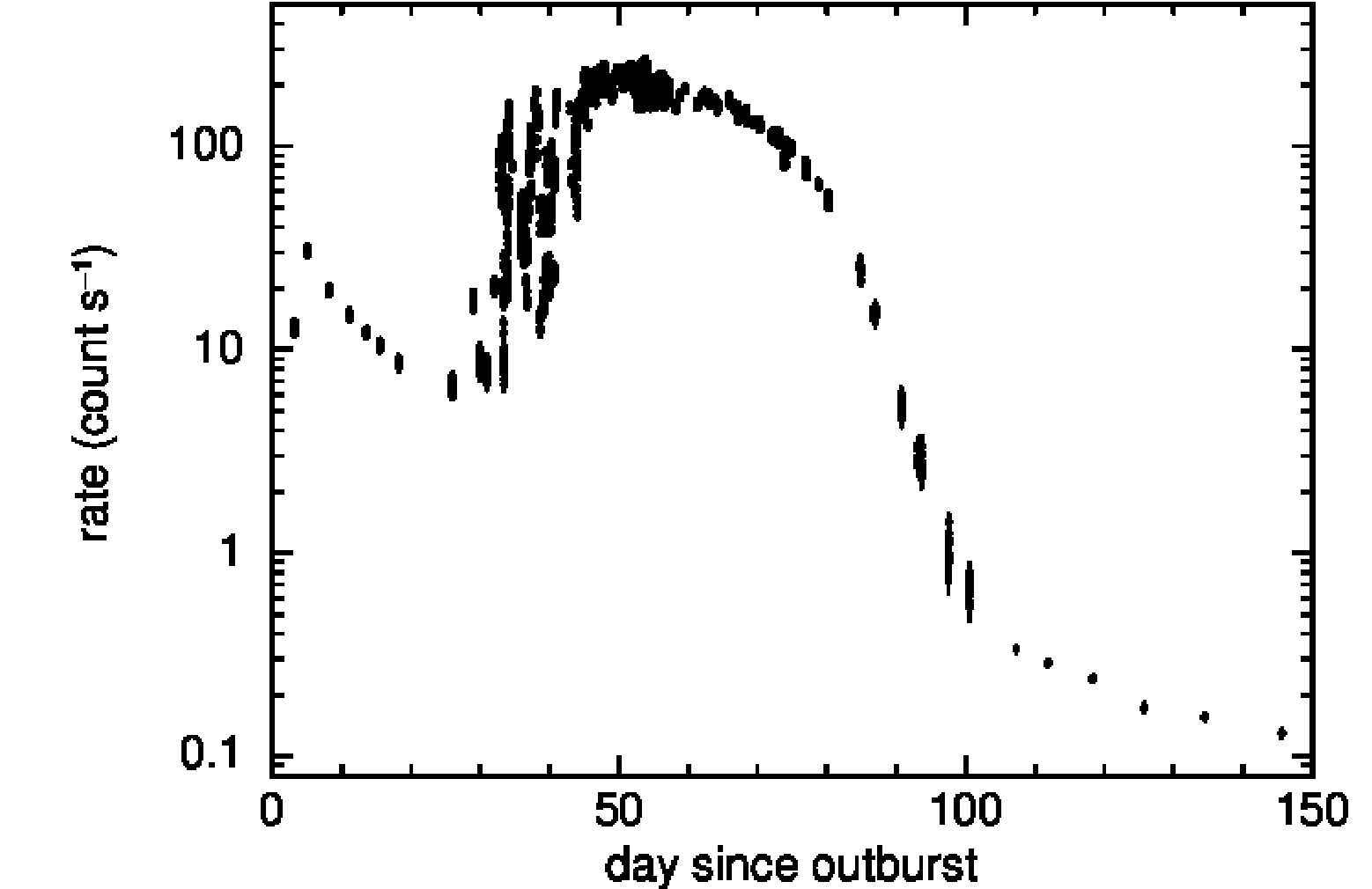}
\end{center}
\caption{Left: \emph{ROSAT} light-curve (0.1--2.4 keV) of V1974 Cyg \cite[taken from][]{kr96}, demonstrating the most detailed X-ray light-curve of a nova obtained before the launch of \emph{Swift}. Right: The X-ray light-curve (0.3--10 keV) of RS Oph \citep{jpo11a} obtained from the first detailed \emph{Swift} monitoring campaign of a nova. Only the first 150 days after outburst are shown.}
\label{comp}\end{figure}

Before the launch of the \emph{Swift Gamma-Ray Burst Explorer Mission}\footnote{In 2018 the satellite was renamed the \emph{Neil Gehrels Swift Observatory} in honour of the former Principal Investigator: https://www.nasa.gov/feature/goddard/2018/nasas-newly-renamed-swift-mission-spies-a-comet-slowdown} \citep{geh04}, only a small number of novae had been detected in the X-ray band. Nova Cyg 1992 (V1974 Cyg) was the most well monitored of these sources, with 18 \emph{ROSAT-PSPC} observations collected \citep{kr96}. As the left-hand panel of Fig.~\ref{comp} shows, the nova was found to brighten, becoming a super-soft source and plateauing in X-rays for a few hundred days, before fading away
rapidly. The shape of the light-curve was explained as the unveiling of the
X-ray source as the ejecta cleared, with the source turning off as
nuclear burning ceased \citep{kr96}.

Around a decade after these observations, on 2004 November 20, \emph{Swift} was launched: a mission designed to detect and follow-up Gamma-Ray Bursts (GRBs), but also very well-suited to monitor transient sources such as novae. In 2006 February, the recurrent nova RS Oph went into outburst. Thanks to the daily planning of \emph{Swift} observing
schedules, monitoring of the nova began within three days of the outburst; the most recent data point was actually collected in 2014 May, more than 3000 days later. Fig.~\ref{comp} (right-hand panel) plots the X-ray data collected, demonstrating that, while the underlying shape is similar to
V1974 Cyg - an underlying rise, plateau and fall is visible - the detailed monitoring revealed deviations from the relatively smooth time-series seen in the earlier nova. In particular, the rise to peak count rate was not at all monotonic, but rather showed high amplitude variability -- an order of magnitude or more in about 12 hours. 

This high-cadence monitoring campaign of RS~Oph \citep[details published by][for example]{bode06c, jpo11a, vay11}  clearly highlighted the abilities of \emph{Swift} in this field, inspiring many more observations of novae by the mission. 

A number of \emph{Swift} nova synopsis papers have been compiled over the years \citep{ness07a, schwarz11, jpo15}. In this article we add to these results, presenting the most complete sample of \emph{Swift}-monitored novae to date, concentrating mainly on the SSS emission.

\section{Swift observations of supersoft novae}

Between 2006 and the end of 2017, \emph{Swift} detected 30 novae in either our Galaxy or the Magellanic Clouds\footnote{We do not consider the many M31 novae here.} with SSS emission. In addition, more than 50 other Galactic novae were observed but not detected in the X-ray band, or only showed hard emission; we do not include these objects in this paper.
We have, however, included V2362~Cyg and V1534~Sco in this current sample despite there being no obvious detection of SSS emission in these datasets, because a substantial amount of \emph{Swift} data was collected. Observations of V2362~Cyg did not begin for almost 200 days after outburst, while in the case of V1534~Sco, the softening of the spectrum can be modelled by a decrease in the absorbing column which remains relatively high (10$^{23-21}$~cm$^{-2}$) throughout the observations \citep{page14e}.

 Of these 32 novae, most were monitored with both the X-ray Telescope \citep[XRT;][]{burr05} and the UV/Optical Telescope \citep[UVOT;][]{rom05}, though in some cases the UV source was too bright for conventional photometry. Table~\ref{table:outburst} lists these novae, together with their outburst times from the literature. For confirmed recurrent novae, marked with an asterix, only the most recent outburst date (i.e., that observed by \emph{Swift}) is listed. V959 Mon was discovered as a new GeV source by LAT at a time when that area of the sky was too close to the Sun for ground-based (or \emph{Swift}) observations; the optical nova was confirmed some seven weeks later \citep{cheung12, fuji12}. Nova SMC 2012 was announced four months after the actual outburst, when a new transient detection system was implemented by OGLE \citep[Optical Gravitational Lensing Experiment; ][]{wyrz12}.

\begin{sidewaystable*}
\caption{Outburst information from the literature for the novae presented in this paper. The last column gives the reference for the date of outburst. $^{*}$Recurrent nova. Date is most recent outburst.  $^{1}$While the optical nova was discovered on 2012-08-09.8061, the $\gamma$-ray source was found almost 2 months earlier. $^{2}$Outburst occurred between 2012-02-24 and 2012-06-05. }
\begin{tabular}{llll}
\hline
Nova & Alternative names & Date of outburst  & Reference  \\
 & & UT\\
\hline
\hline
RS Oph$^{*}$ & --- & 2006-02-12.8 & \cite{nar06}\\
\hline
V2362 Cyg & Nova Cyg 2006 & 2006-04-02.807 & \cite{nak06}\\
\hline
V1280 Sco & Nova Sco 2007 & 2007-02-04.8624 & \cite{yam07a}\\
\hline
V1281 Sco & Nova Sco 2007 No. 2 & 2007-02-19.8593 & \cite{yam07b}\\ 
\hline
V458 Vul & Nova Vul 2007 & 2007-08-08.54 & \cite{nak07}\\
\hline
V598 Pup & XMMSL1 J070542.7-381442 & 2007-10-09 & \cite{read07,read08}\\
\hline
V597 Pup & Nova Pup 2007 & 2007-11-14.23 & \cite{per07}\\
\hline
V2468 Cyg & Nova Cyg 2008a & 2008-03-07.801 & \cite{nak08a}\\
\hline
V2491 Cyg & Nova Cyg 2008 No. 2 & 2008-04-10.728 & \cite{nak08b}\\
\hline
HV Cet & CSS 081007:030559+054715 & 2008-10-07.381 & \cite{apb12}\\
\hline
Nova LMC 2009a$^{*}$  & Nova LMC 1971b & 2009-02-05.067 & \cite{lil09}\\
\hline
V1213 Cen & Nova Cen 2009 & 2009-05-08.235 & \cite{poj09}\\
\hline
V2672 Oph & Nova Oph 2009 & 2009-08-16.515 & \cite{nak09}\\
\hline
KT Eri & Nova Eri 2009 & 2009-11-14.632 & \cite{yam09}\\
\hline
U Sco$^{*}$ & --- & 2010-01-28.4385 & \cite{sch10} \\
\hline
V407 Cyg & --- & 2010-03-10.797 & \cite{nish10} \\
\hline
T Pyx$^{*}$  & --- & 2011-04-14.2391 & \cite{waagen11}\\
\hline
Nova LMC 2012 & TCP J04550000-7027150 & 2012-03-26.397 & \cite{seach12} \\
\hline
V5589 Sgr & Nova Sgr 2012; PNV J17452791-2305213 & 2012-04-21.011 & \cite{kor12} \\
\hline
V959 Mon & Nova Mon 2012; PNV J06393874+0553520 & 2012-06-22$^{1}$ & \cite{fuji12}\\
& & &\cite{cheung12} \\
\hline
Nova SMC 2012 & OGLE-2012-NOVA-002 & 2012-06-05$^{2}$ & \cite{wyrz12}  \\
\hline
V339 Del & Nova Del 2013; PNV J20233073+2046041 & 2013-08-14.584 & \cite{nak13} \\
\hline
V1369 Cen & Nova Cen 2013; PNV J13544700-5909080 & 2013-12-02.692 & \cite{seach13} \\
\hline
V745 Sco$^{*}$ & --- & 2014-02-06.694 & \cite{waagen14} \\
\hline
V1534 Sco & Nova Sco 2014;  TCP J17154683-3128303 & 2014-03-26.84867 & \cite{nish14}\\
\hline
V1535 Sco & Nova Sco 2015; PNV J17032620-3504140 & 2015-02-11.8367 & \cite{nak15}\\
\hline
V5668 Sgr & Nova Sgr 2015 No. 2; PNV J18365700-2855420 & 2015-03-15.634 & \cite{seach15} \\
\hline
Nova LMC 1968-12a$^{*}$  & OGLE-2016-NOVA-01 & 2016-01-21.2094 & \cite{mroz16a}\\
\hline
V407 Lup & Nova Lup 2016; ASASSN-16kt & 2016-09-24.00  & \cite{stan16} \\
\hline
Nova SMC 2016 & --- & 2016-10-09-09.2   & \cite{mroz16b}\\
\hline
V5855 Sgr & Nova Sgr 2016 No. 3; TCP J18102829-2729590 & 2016-10-20.383 & \cite{nak16}\\
\hline
V549 Vel & Nova Vel 2017; ASASSN-17mt & 2017-09-24.39 & \cite{stan17} \\
\end{tabular}
\label{table:outburst}
\end{sidewaystable*}

\begin{figure}
  \begin{center}
    \caption{Novae from 2006. The hardness ratio bands vary between different novae, as explained in the text.}\label{2006}
    \includegraphics*[width=8cm,angle=-90]{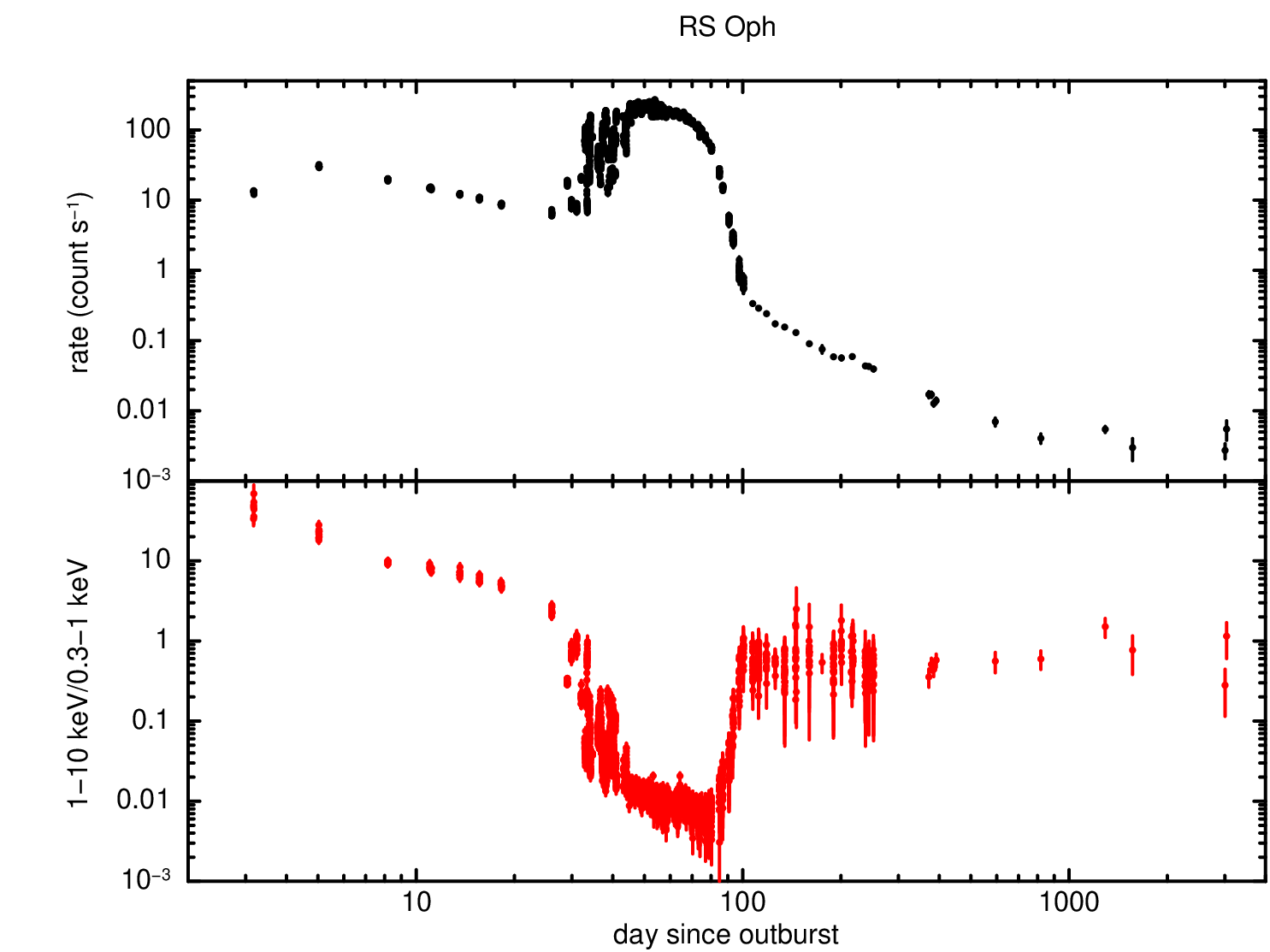}
    \includegraphics*[width=8cm,angle=-90]{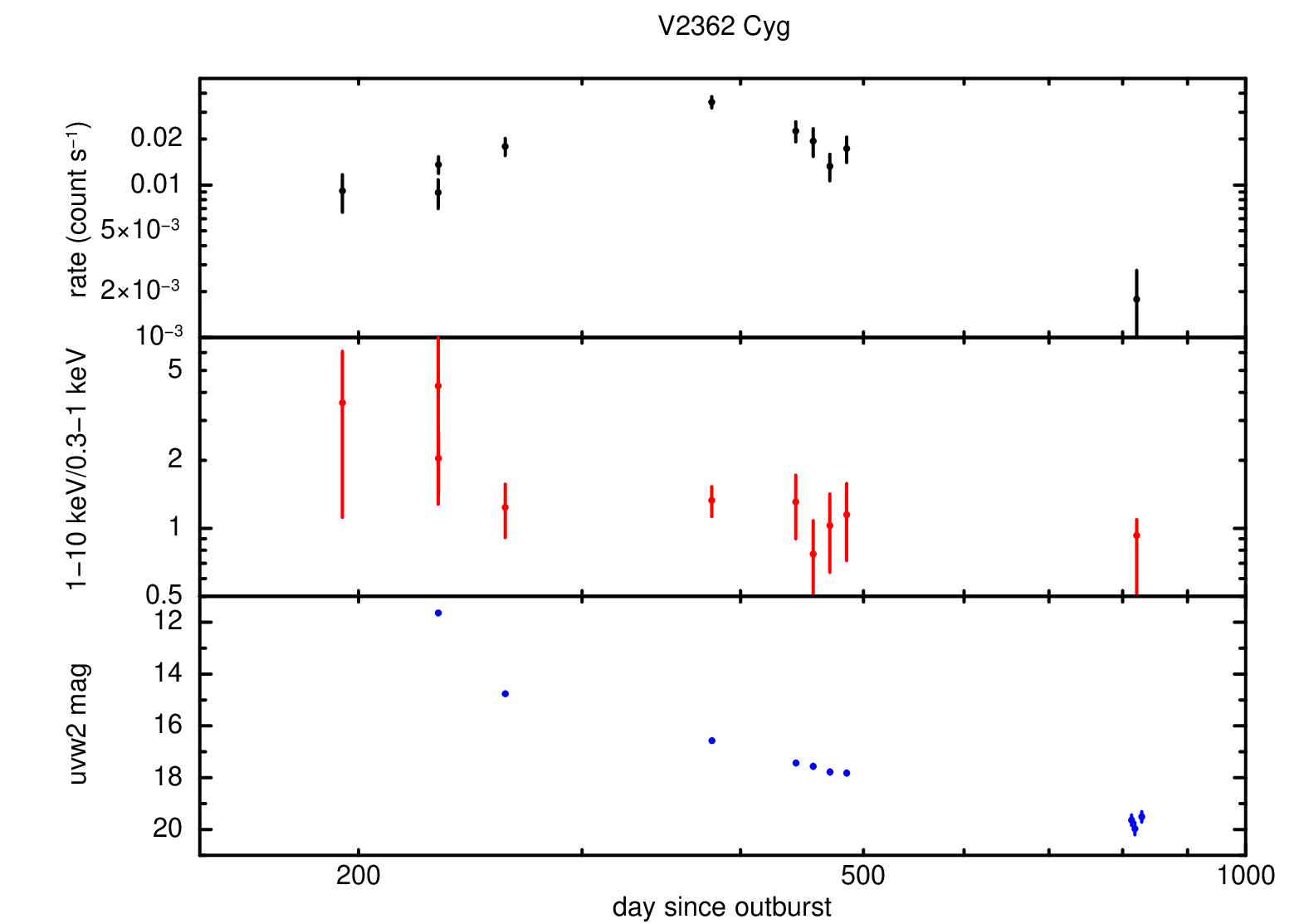}
\end{center}

\end{figure}

\begin{figure}
  \begin{center}
  \caption{Novae from 2010. The hardness ratio bands vary between different novae, as explained in the text.} \label{2010} 
  \includegraphics*[width=8cm,angle=-90]{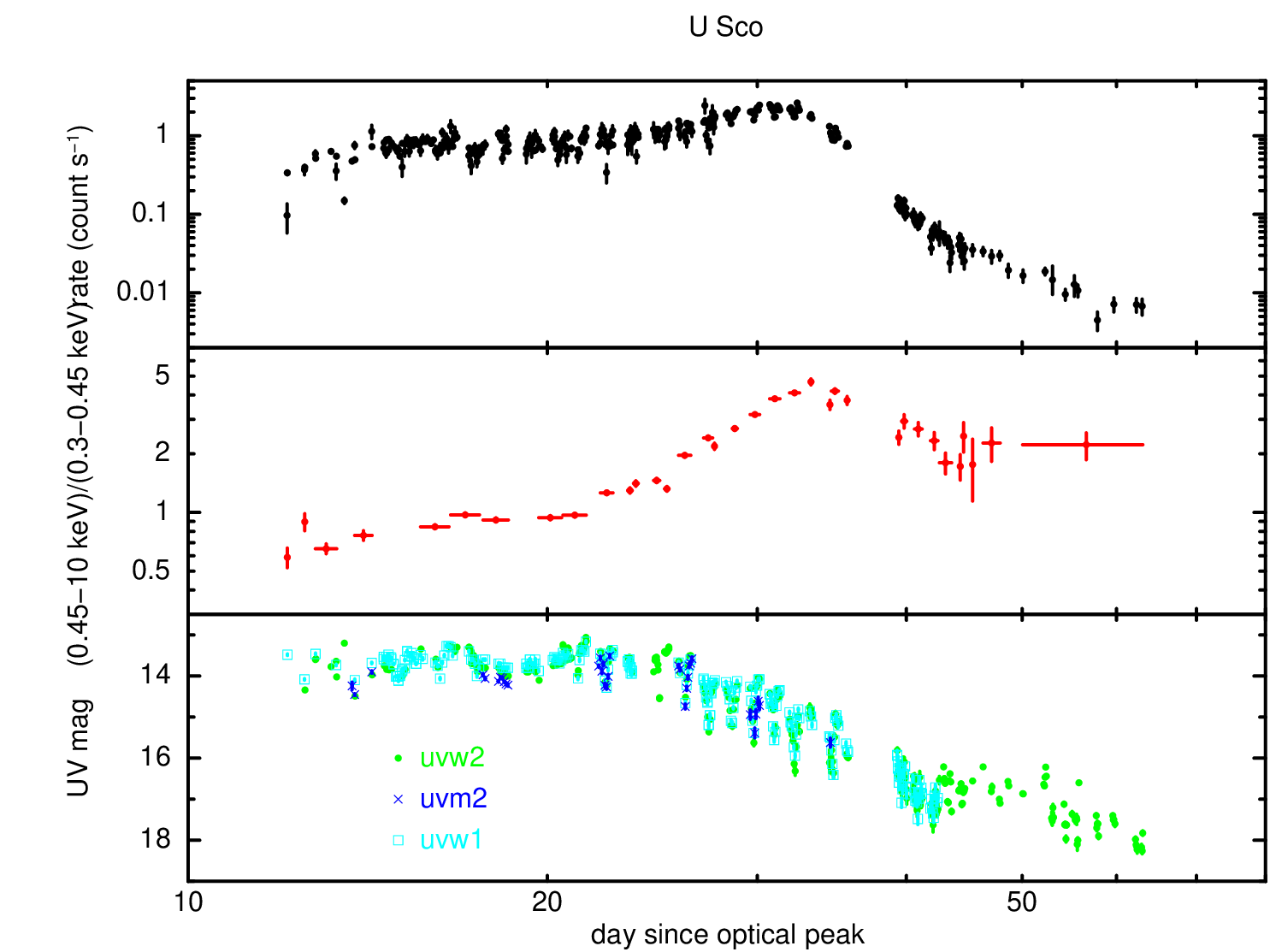}\\
    \includegraphics*[width=8cm,angle=-90]{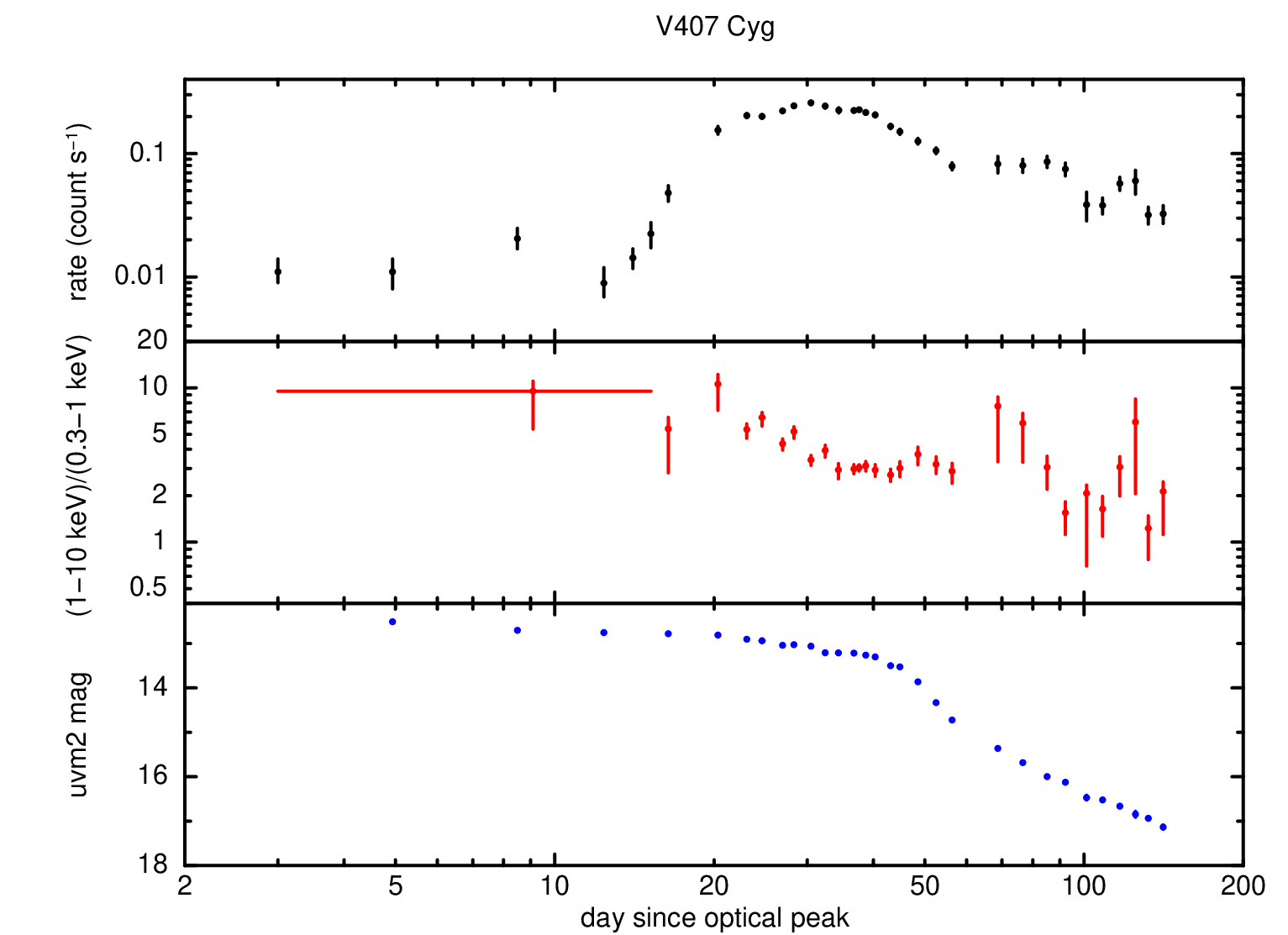}
\end{center}
\end{figure}

\begin{figure}
  \begin{center}
  \caption{Nova from 2011. The hardness ratio bands vary between different novae, as explained in the text.}\label{2011}  
    \includegraphics*[width=8cm,angle=-90]{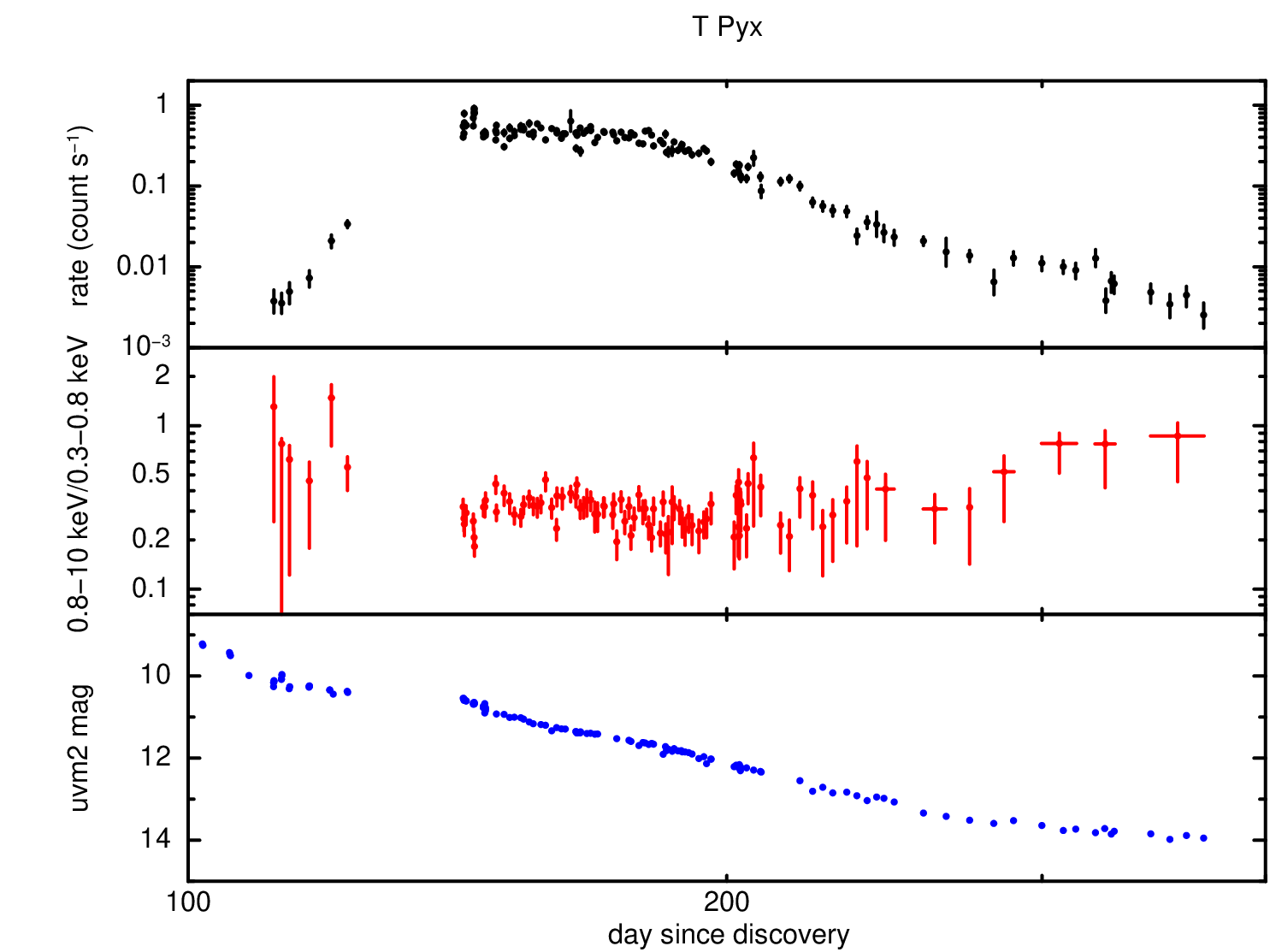}
\end{center}

\end{figure}

\begin{figure}
  \begin{center}
    \caption{Novae from 2013. The hardness ratio bands vary between different novae, as explained in the text.}\label{2013}
  \includegraphics*[width=8cm,angle=-90]{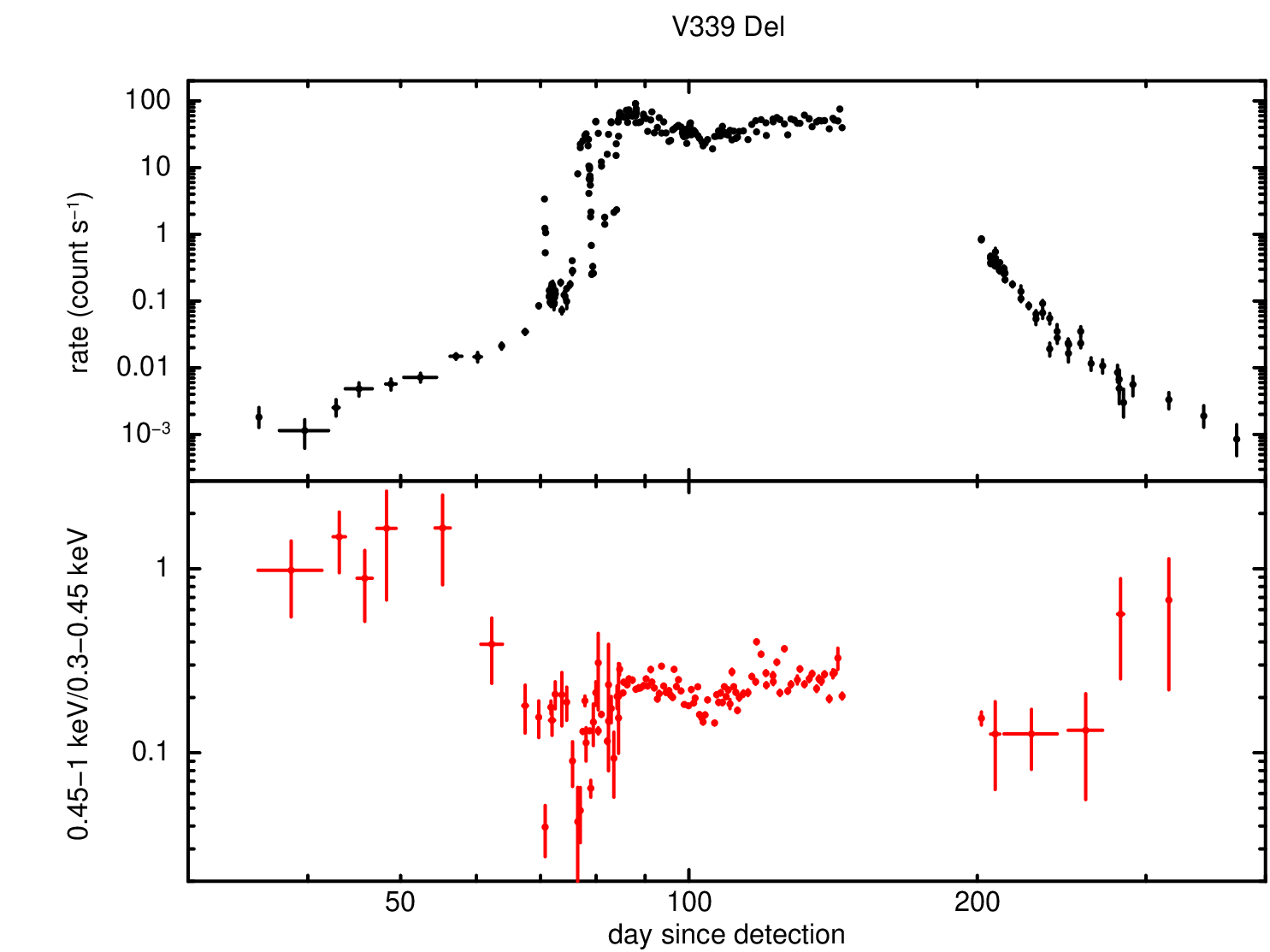}\\
  \includegraphics*[width=8cm,angle=-90]{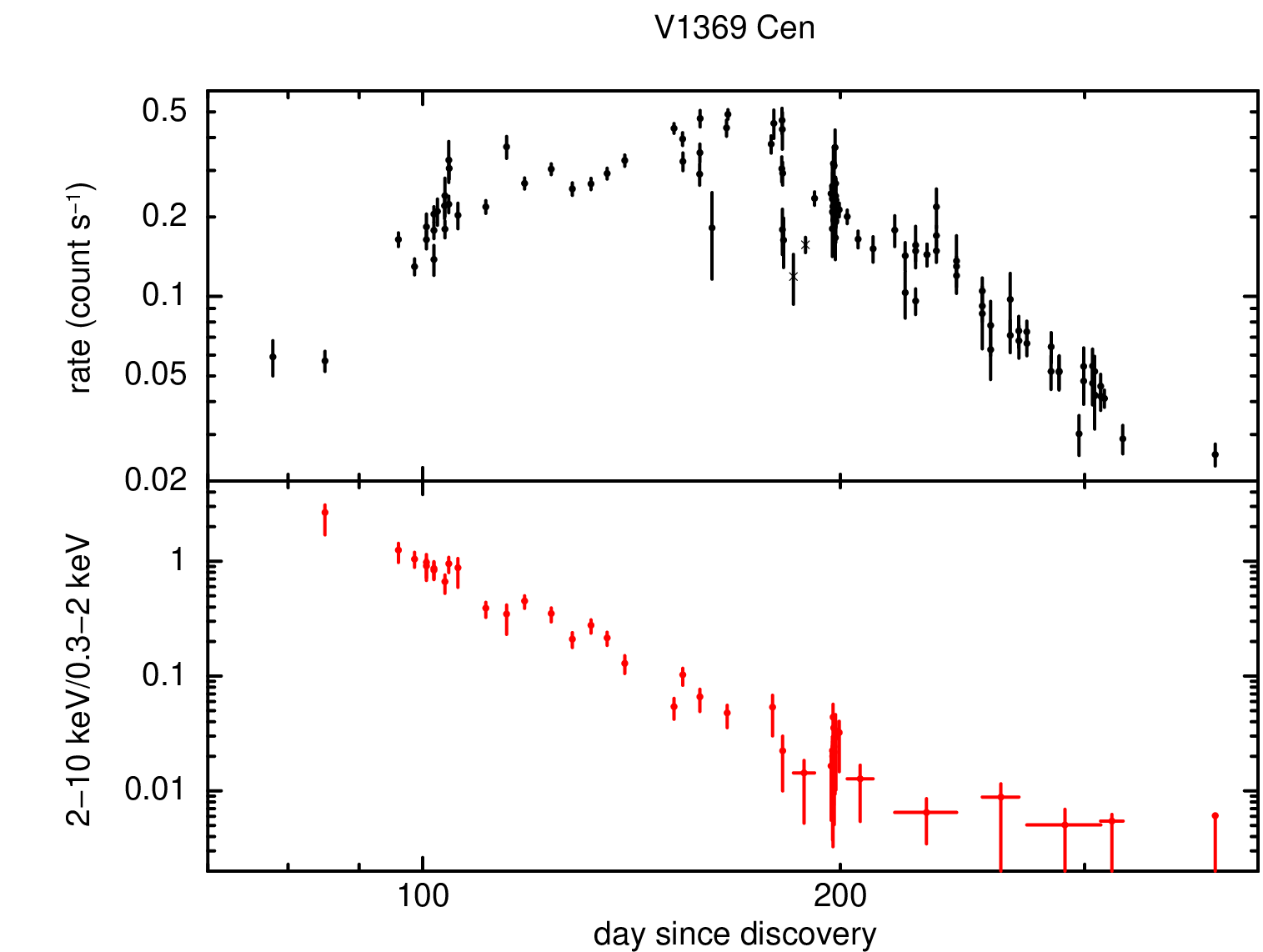}\\    
\end{center}

\end{figure}

\begin{figure}
  \begin{center}
    \caption{Novae from 2014. The hardness ratio bands vary between different novae, as explained in the text.}\label{2014}
   \includegraphics*[width=8cm,angle=-90]{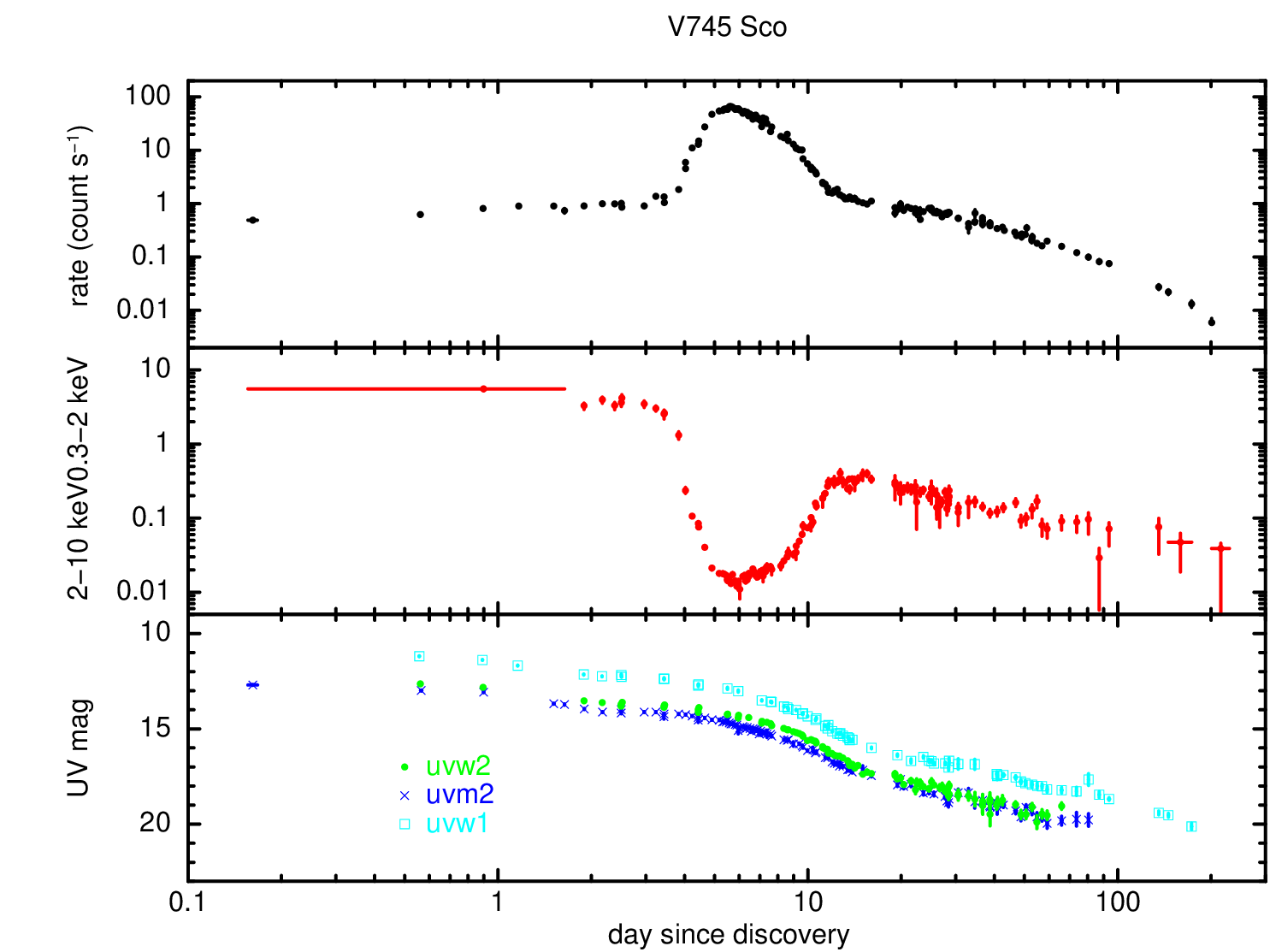}\\
  \includegraphics*[width=8cm,angle=-90]{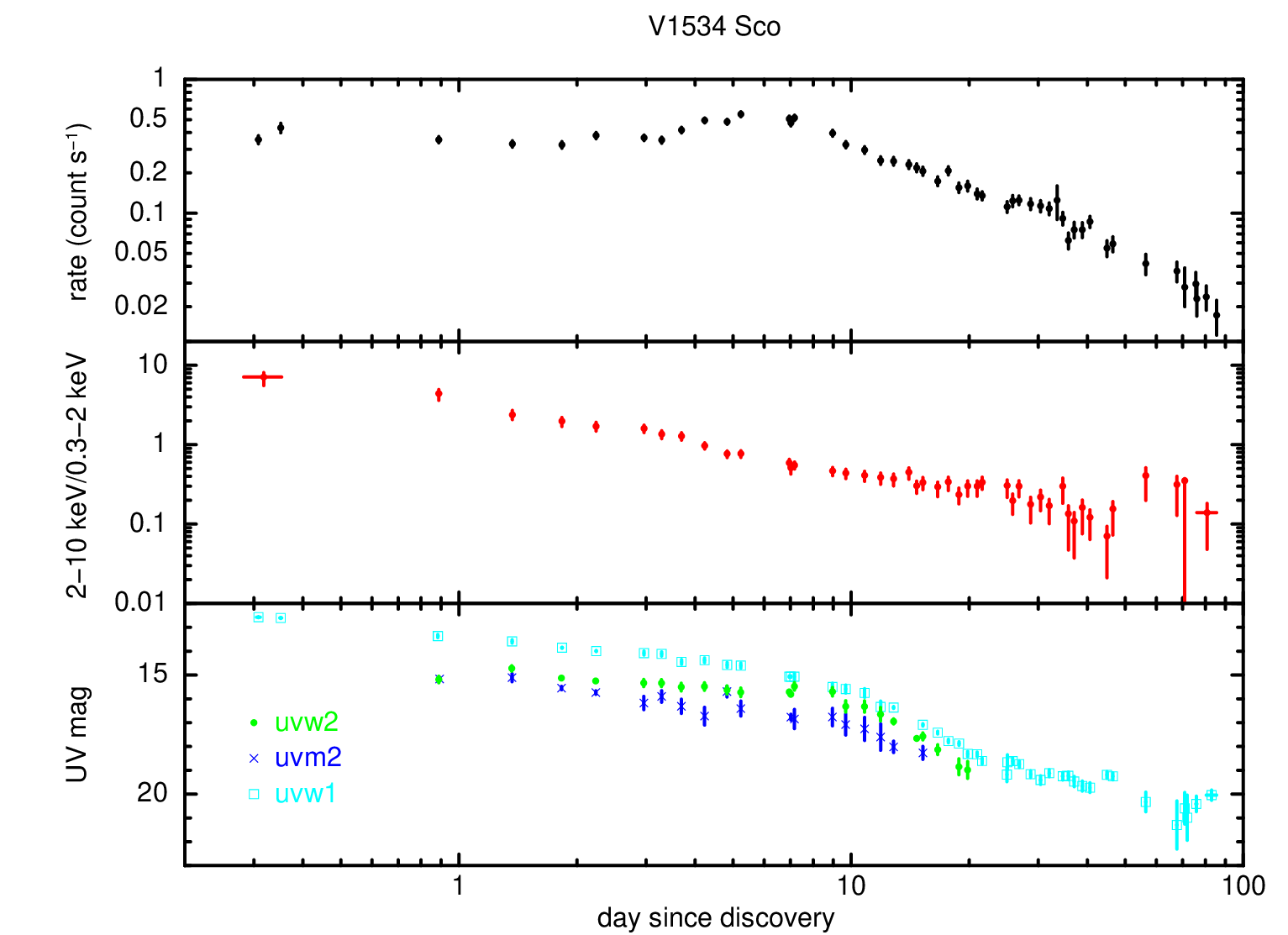}\\
\end{center}

\end{figure}

\begin{figure}
  \begin{center}
    \caption{Novae from 2015. The hardness ratio bands vary between different novae, as explained in the text.}\label{2015}
   \includegraphics*[width=8cm,angle=-90]{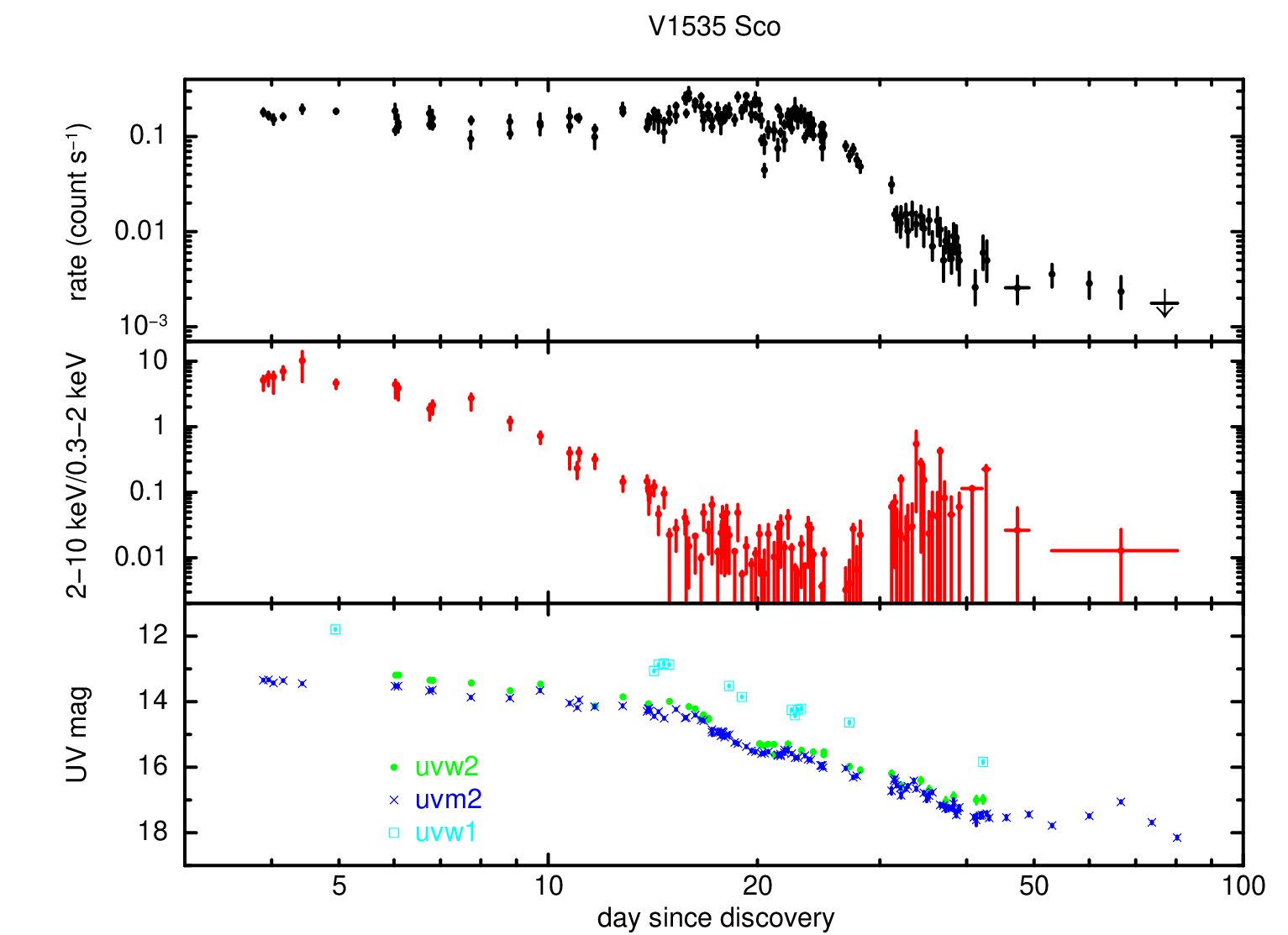}\\
  \includegraphics*[width=8cm,angle=-90]{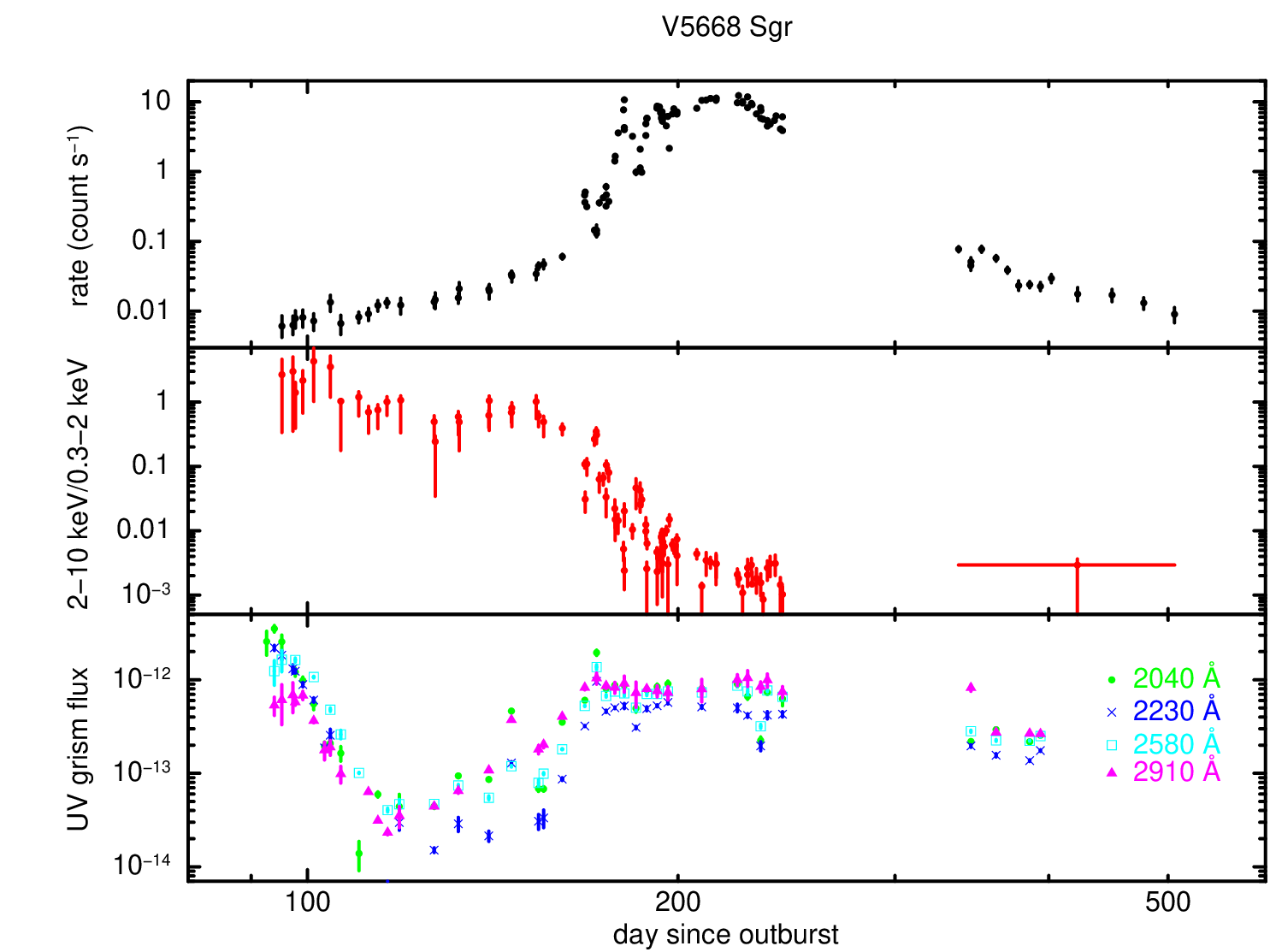}   
\end{center}

\end{figure}

\begin{figure}
  \begin{center}
  \caption{Nova from 2017. The hardness ratio bands vary between different novae, as explained in the text.}  \label{2017}
\includegraphics*[width=8cm,angle=-90]{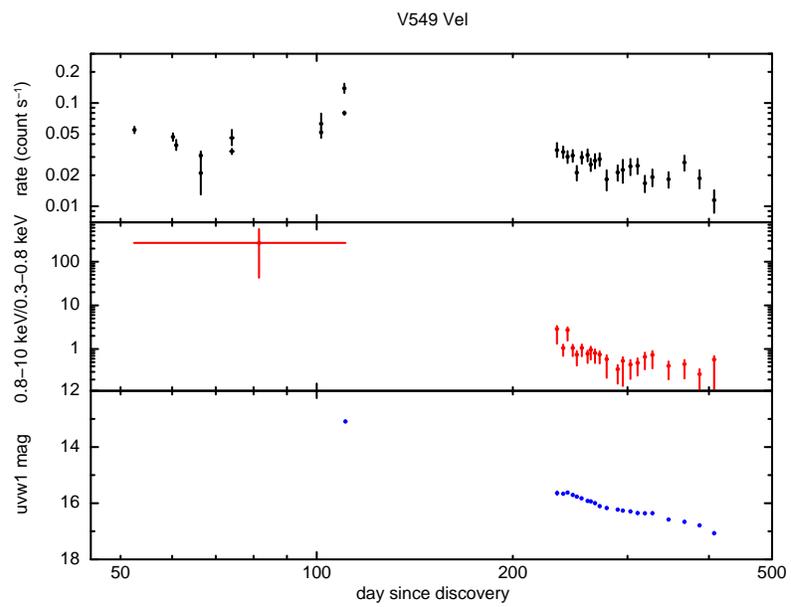}
\end{center}
\end{figure}

Figures~\ref{2006}--\ref{2017} show the \emph{Swift} results for these novae\footnote{with the exception of V5855~Sgr, since there was only a single (soft) X-ray detection obtained for this source}. In each case, the top panel shows the XRT light-curve over 0.3--10~keV, while the X-ray hardness ratio is plotted in the second panel, here defined as hard-band counts divided by soft-band counts. The precise cut between the soft and hard bands has been chosen on a case-by-case basis, depending on the shape of the X-ray spectrum. In most cases this cut is taken to separate the SSS emisison from the harder (shock) component. However, for some novae the only emission clearly detected is soft (i.e. the vast majority of counts lie below about 1~keV); in these cases (HV Cet, V339 Del, V407 Lup and Nova SMC 2016), the hardness ratio compares two bands within the soft emission. When standard UVOT photometry could be utilised, a third panel shows the magnitude light-curve in whichever filters were predominantly used. In the cases of KT Eri and V5668 Sgr, all the UVOT observations were obtained using the grism; the panel therefore shows flux estimated from the grism spectra. In order to display the main emission intervals more clearly, occasionally early or late data points have been excluded from the plots. 

\begin{longfigure}{c}
  \caption{Novae from 2007. The hardness ratio bands vary between different novae, as explained in the text. Note that the light-curve of V598 Pup only covers a short time interval, and is therefore plotted on a linear time axis.}\label{2007}\\
  \endLFfirsthead
  \caption{Novae from 2007 -- continued from previous page}\\
  \endLFhead
  \endLFfoot
  \endLFlastfoot
      \includegraphics*[width=8cm,angle=-90]{V1280Sco_lc-hr-uv_new.ps}\\
  \includegraphics*[width=8cm,angle=-90]{V1281Sco_lc-hr-uv_zoom.ps}\\
  \includegraphics*[width=8cm,angle=-90]{V458Vul_lc-hr-uv_zoom.ps}\\
  \includegraphics*[width=8cm,angle=-90]{V598Pup_lc-hr-uv_zoom_new.ps}\\
  \includegraphics*[width=8cm,angle=-90]{V597Pup_lc-hr-uv_zoom.ps}\\
\end{longfigure}

\begin{longfigure}{c}
  \caption{Novae from 2008. The hardness ratio bands vary between different novae, as explained in the text.}\label{2008}\\
\endLFfirsthead
  \caption{Novae from 2008 -- continued from previous page}\\
  \endLFhead
  \endLFfoot
  \endLFlastfoot
    \includegraphics*[width=8cm,angle=-90]{V2468Cyg_lc-hr-uv_new.ps}\\
    \includegraphics*[width=8cm,angle=-90]{V2491Cyg_lc-hr-uv.ps}\\
    \includegraphics*[width=8cm,angle=-90]{HVCet_lc-hr-uv.ps}\\

\end{longfigure}

\begin{longfigure}{c}
  \caption{Novae from 2009. The hardness ratio bands vary between different novae, as explained in the text.}\label{2009}\\
\endLFfirsthead
  \caption{Novae from 2009 -- continued from previous page}\\
  \endLFhead
  \endLFfoot
  \endLFlastfoot
    \includegraphics*[width=8cm,angle=-90]{NovaLMC2009a_lc-hr-uv_zoom.ps}\\
    \includegraphics*[width=8cm,angle=-90]{V1213Cen_lc-hr-uv.ps}\\
    \includegraphics*[width=8cm,angle=-90]{V2672Oph_lc-hr-uv.ps}\\
    \includegraphics*[width=8cm,angle=-90]{KTEri_lc-hr-uv.ps}\\

\end{longfigure}

\begin{longfigure}{c}
  \caption{Novae from 2012. The hardness ratio bands vary between different novae, as explained in the text.}\label{2012}\\
\endLFfirsthead
  \caption{Novae from 2012 -- continued from previous page}\\
  \endLFhead
  \endLFfoot
  \endLFlastfoot
    \includegraphics*[width=8cm,angle=-90]{NovaLMC2012_lc-hr-uv_zoom.ps}\\
    \includegraphics*[width=8cm,angle=-90]{NovaSgr2012_lc-hr-uv.ps}\\
    \includegraphics*[width=8cm,angle=-90]{V959Mon_lc-hr-uv_new.ps}\\
    \includegraphics*[width=8cm,angle=-90]{NovaSMC2012_lc-hr-uv_zoom_new.ps}

\end{longfigure}

\begin{longfigure}{c}
  \caption{Novae from 2016. The hardness ratio bands vary between different novae, as explained in the text.}\label{2016}\\
\endLFfirsthead
  \caption{Novae from 2016 -- continued from previous page}\\
  \endLFhead
  \endLFfoot
  \endLFlastfoot
 \includegraphics*[width=8cm,angle=-90]{LMC1968_lc-hr-uv.ps}\\
  \includegraphics*[width=8cm,angle=-90]{V407Lup_lc-hr-uv.ps}\\
  \includegraphics*[width=8cm,angle=-90]{SMC2016_lc-hr-uv.ps}\\

\end{longfigure}

Table~\ref{table:xrt} lists the \emph{Swift} timings for the novae: the range of first to last observations taken, the first X-ray detection (hard or soft) and the first detection of supersoft emission. In the instances where the observations cover an extended period, there were long gaps between the later snapshots of data. For a couple of the novae (marked in the table), there were also long gaps between the initial and subsequent observations.

\begin{landscape}
\begin{longtable}{lllll}  
  \caption{{\emph Swift} results for the novae presented in this paper. $^{1}$No observations taken between days 5.3 and 822.5, or between 928.1 and 2285.8. $^{2}$No observations between days 1.9 and 337.7. $^{3}$No observations between days 14.6 and 323.0.  } \label{table:xrt}\\

\hline
Nova & Interval of \emph{Swift} obs.  & First XRT detection & First SSS detection & References \\
&(day after outburst)&(day after outburst)&(day after outburst)\\
\hline
\hline
\endfirsthead

\multicolumn{5}{c}%
{{\bfseries \tablename\ \thetable{} -- continued from previous page}} \\
\hline
Nova & Interval of \emph{Swift} obs.  & First XRT detection & First SSS detection & References \\
&(day after outburst)&(day after outburst)&(day after outburst)\\
\hline
\hline
\endhead

\hline \multicolumn{5}{|r|}{{Continued on next page}} \\ \hline
\endfoot

\hline \hline
\endlastfoot

RS Oph & 3.2--3046.9 & 3.2 & 26.0 &\cite{jpo06a,jpo06b, jpo06c, jpo06d}\\
& & & & \cite{bode06a, bode06b, bode06c}\\
& & & & \cite{jpo11a}\\
\hline
V2362 Cyg & 194.3--820.7 & 194.3 & --- & \cite{lynch08}\\
\hline
V1280 Sco & 5.3--4401.1$^{1}$ & 822.5 & 2285.8 & \cite{jpo07}\\
 & & & & \cite{ness09a}\\
\hline
V1281 Sco & 1.9--819.0$^{2}$ & 337.7 & 337.7 & \cite{jpo07}\\
& & & &\cite{ness08a}\\
\hline
V458 Vul & 1.1--1487.9 & 70.5 & 397.6  & \cite{drake07} \\
& & & &\cite{drake08a}\\
& & & &\cite{drake08b}\\
& & & &\cite{ness09b}\\
\hline
V598 Pup & 168.8--255.8 & 168.8 & 168.8 & \cite{read08}\\
& & & &\cite{page09}\\
\hline
V597 Pup & 6.1--584.6 & 55.0 & 120.3 & \cite{ness08b}\\
\hline
V2468 Cyg & 458.4--2311.6 & 458.4 & 1266.7 & \cite{schwarz09a}\\
& & & &\cite{schwarz09b}\\
& & & &\cite{page12b}\\
\hline
V2491 Cyg & 1.0--236.0 & 1.0 & 33.3 & \cite{ib08}\\
& & & &\cite{ibet08}\\
& & & &\cite{kuulk08}\\
& & & &\cite{page08,page10}\\
& & & &\cite{jpo08}\\
& & & &\cite{ib09}\\
& & & &\cite{ness11}\\
& & & &\cite{tak11}\\
\hline
HV Cet & 34.1--299.6 & 34.1 & 34.1 & \cite{schwarz08}\\
& & & &\cite{apb08a,apb12}\\
& & & &\cite{jpo09}\\
\hline
Nova LMC 2009a & 9.2--382.3 & 63.0 & 70.7 & \cite{bode09a,bode09b, bode16}\\
\hline
V1213 Cen & 14.6--1407.1$^{3}$ & 323.0 & 323.0 & \cite{schwarz10}\\
\hline
V2672 Oph & 1.5--72.5 & 1.5 & 14.7 & \cite{schwarz09c}\\
& & & &\cite{tak14}\\
\hline
KT Eri & 13.1--734.3 & 39.9 & 55.5 & \cite{bode10,bode16}\\
& & & &\cite{apb10}\\
\hline
U Sco &  0.6--63.1 & 2.6 & 12.1 & \cite{schlegel10a,schlegel10b}\\
& & & &\cite{jpo10}\\
& & & &\cite{pag15}\\
& & & &\cite{tak13}\\
\hline
V407 Cyg & 3.0--1228.1 & 3.0 & 12.4 & \cite{shore11}\\
& & & &\cite{nel12a}\\
\hline
T Pyx & 0.31--718.9 & 0.31 & 122.8& \cite{kuulk11}\\
 & & & &\cite{jpo11b}\\
& & & &\cite{toff13}\\
& & & &\cite{chom14a}\\
\hline
Nova LMC 2012 & 1.3--671.0 & 18.1 & 18.1 & \cite{page12a}\\
& & & & \cite{schwarz15}\\
\hline
V5589 Sgr & 0.9--108.4 & 19.7 & 64.6 & \cite{sok12}\\
& & & &\cite{nel12b}\\
& & & &\cite{weston16b}\\
\hline
V959 Mon & 58.5--1002.8  &  58.5 & 149.7 & \cite{nel12c, nel12d}\\
& & & &\cite{jpo13a}\\
 & & & &\cite{page13a,page13f}\\
\hline
Nova SMC 2012 & 135.8--439.0 & 135.8 & 270.4 & \cite{schwarz12}\\
& & & &\cite{page13b,page13c}\\
\hline
V339 Del & 0.4--408.0 & 35.6 & 59.7 & \cite{kuulk13a}\\
& & & &\cite{nel13}\\
& & & &\cite{page13d,page13e,page14d}\\
& & & &\cite{pagebeard13}\\
& & & &\cite{jpo13b}\\
& & & &\cite{beard13}\\
& & & &\cite{shore16} \\
\hline
V1369 Cen & 1.2--1967.3 & 78.0 & 128.2 & \cite{kuulk13b}\\
& & & &\cite{page14c}\\
& & & &\cite{mason18}\\
\hline
V745 Sco & 0.2--229.2 & 0.2 & 3.4 & \cite{muk14}\\
& & & &\cite{page14a,page14b,page15d}\\
& & & &\cite{apb14}\\
\hline
V1534 Sco & 0.3--85.5 & 0.3 & --- & \cite{kuulk14}\\
& & & &\cite{page14e}\\
\hline
V1535 Sco & 3.9--80.4 & 3.9 & 11.7 & \cite{nel15}\\
& & & &\cite{lin17}\\
\hline
V5668 Sgr & 2.6--848.2 & 95.3 & 167.9 & \cite{page15a,page15b,page15c}\\
& & & &\cite{gehrz18}\\
\hline
Nova LMC 1968-12a & 2.1--85.9 & 2.1 & 6.4 & \cite{darn16}\\
& & & &\cite{page16a}\\
& & & &Kuin et al. in prep.\\
\hline
V407 Lup & 2.9--683.7 & 150.1 & 150.1 & \cite{apb17a}\\
& & & &\cite{orio17}\\
& & & &\cite{apb17b}\\
& & & & \cite{aydi18b}\\
\hline
Nova SMC 2016 & 5.5--318.7 & 28.2 & 28.2 & \cite{kuin16}\\
& & & &\cite{page16b}\\
 & & & &\cite{aydi18a}\\
\hline
V5855 Sgr & 7.6--129.9 & 129.9 & 129.9 & \cite{nel19} \\

\hline
V549 Vel & 52.5--407.4 & 52.5 & 233.8 & \cite{page18}\\

\end{longtable}
\end{landscape}

\subsection{Highlights of Swift results}

\emph{Swift} observations of SSS emission from novae have discovered a number of unexpected and interesting features, which are briefly discussed below.

\subsubsection{High-amplitude flux variability}

Observations of RS~Oph first identified that the initial rise to peak soft X-ray emission was chaotic. The X-ray count rate was found to vary significantly, sometimes by more than an order of magnitude in 12 hours (Figs.~\ref{comp} and \ref{2006}), before settling at a consistently high count rate for around 20 days, and then fading rapidly away. The emission is typically softer when brighter, both on long and short timescale, though, as discussed in \cite{jpo11b}, occasional counter-examples were found. Monitoring campaigns since RS~Oph have shown this high-amplitude flux variability not to be unique to that source, with other novae showing a similar phenomenon (e.g., KT~Eri, Nova~LMC~2009a, V2491~Cyg, V458~Vul; see references in Table~\ref{table:xrt}). However, not all novae followed in detail show this variability: V745~Sco \citep{page15d} is a prime example of a smooth rise to peak SSS emission.

As has been previously discussed  \citep[e.g., ][]{pageconf13, schwarz11, jpo11a}, these large scale changes in flux are at least partly due to variable visibility of the hot WD. Following a nova explosion, material will be expelled from the WD surface. If these clumpy ejecta pass through the observer's line of sight, then the count rate will drop as the X-rays are absorbed. Fits to the SSS X-ray spectra also reveal variations in the photospheric temperature \citep[e.g., ][]{pageconf14, jpo11a}, where a lower temperature (below $\sim$~30~eV) can lead to part, if not most, of the SSS emission being below the XRT bandpass, thus decreasing the measured count rate. 


\subsubsection{Quasi-periodic oscillations}

Continuing its surprises, RS~Oph X-ray data also demonstrated a 35s quasi-periodic oscillation \citep[QPO;][]{jpo11b, jpo06b, apb08b}; this periodicity was confirmed in \emph{XMM-Newton} data by \cite{ness07b}. Subsequently QPOs were also identified in other bright, supersoft novae, for example KT~Eri \citep{apb10}, V339~Del \citep{beard13}, V745~Sco \citep{apb14} and V5668~Sgr \citep{page15c}. The persistent SSS Cal 83 also shows a similar periodicity \citep{odendaal14, ness15b}. \cite{ness15b} is a summary paper of QPO results found in \emph{XMM-Newton} and \emph{Chandra} light-curves, while the \emph{Swift} data will be published in Beardmore et al. (in prep.). 

\subsubsection{X-ray and UV variability}

During the evolution of a nova, nuclear burning is expected to continue at constant bolometric luminosity during the phase of stable shell burning \citep[e.g.][]{gal76, mac85}, with the spectral energy distribution shifting to higher energies as the outburst progresses. Such a phase is not always obvious from observational data, though. \cite{page10} analysed the \emph{Swift} X-ray spectra of V2491~Cyg, finding no obvious evidence for a constant bolometic luminosity phase. Likewise, \cite{page15d} investigated the apparent lack of the phase in V745~Sco; forcing the bolometric luminosity of the soft component model to remain constant, the photospheric radius would need to decrease by a factor of $\sim$~30 over this interval. However, \cite{aydi18a, aydi18b} found that the bolometric luminosity appeared close to constant for 100--150~day in both Nova SMC 2016 and V407~Lup, and RS~Oph showed a plateau in luminosity for a few tens of days \citep{jpo11a}.

While some of the X-ray light-curves presented here do show plateau phases during the SSS emission, others are far from constant. The X-ray count rate is certainly not a precise proxy for the bolometric luminosity -- variations in spectral shape and absorption will alter the count rate to luminosity conversion. In addition, the bolometric correction can be quite uncertain for low temperatures peaking towards the lower energy bound of the XRT bandpass.  In summary, the presence of a constant bolometric luminosity phase is not always clear in \emph{Swift}-XRT observations.

As demonstrated in Figs.~\ref{2006}--\ref{2017}, \emph{Swift} data are usually obtained simultaneously with both the XRT and UVOT. Comparison of the variability across these two bands shows different patterns for different novae. HV~Cet (Fig.~\ref{2008}) is an example where the X-ray and UV emission is modulated in phase, with a 1.77-day period. As discussed in \cite{apb12} and \cite{jpo15}, this is thought to be caused by obscuration in a high-inclination system. In the case of HV~Cet, it is believed that the WD itself is permanently hidden by a scattering region, and that the X-rays detected have been scattered into our line of sight, also explaining the sub-Eddington luminosities measured. The UV is then formed through reprocessed X-rays, with the emission occulted by the disc rim
each 1.77 day orbit. \cite{ness13} present a study of SSS grating spectra, finding two distinct types: those dominated by absorption lines (termed SSa) and those where emission lines are most prominent (SSe). They interpret the SSe systems as those where the central source is obscured, and HV~Cet is classified as an SSe source, supporting the description above.

In contrast, for V458~Vul \citep[Fig.~\ref{2007};][]{ness09b,schwarz11} there is an approximate anti-correlation between the X-ray and UV data, similar to the results found for persistent SSS, such as RX~J0513.9$-$6951 \citep{reinsch00}, where the variability is speculated to be caused by changes in the mass accretion rate. As the accretion rate onto the WD increases, its photosphere expands and cools, shifting the peak of the emission into the extreme UV; when the photosphere shrinks back down, the X-rays would become stronger again.

Finally, there are situations where there is no discernible correlation between the X-ray and UV photons. In the cases of V2491~Cyg and V745~Sco \citep[Figs.~\ref{2008} and \ref{2014};][]{page10, ness11, page15d}, for example, the X-rays brighten, peak and decay, while the UV simply fades over time. V5668~Sgr \citep[Fig.~\ref{2015};][]{gehrz18} shows an example of a dust dip and recovery in the UV band which is not observed in the X-ray band. Such examples suggest the X-ray and UV emitting regions are distinct in these novae.

It has previously been noted that recurrent novae often show a plateau in their optical light-curves coincident in time with the SSS phase \citep[][and references therein]{hach06}, which is speculated to arise from the re-radiation of the bright SSS emission from an accretion disc. Such flattenings can also sometimes be seen in UV light-curves, with KT~Eri and U~Sco (Figs.~\ref{2009} and \ref{2010}) being good examples in the current \emph{Swift} sample.

Periodic variations in the UVOT data have been used to identify orbital periods in some novae \citep[for example,][]{apb12, aydi18b, bode16}. Given its orbit of $\sim$~1.5~hr, \emph{Swift} can measure timescales close to a day which are generally difficult to do from the ground. However, the detection of intermediate polar (IP)-like spin periods ($\sim$~300--1000~s) is more difficult with \emph{Swift} data, given its observing strategy (continuous snapshots of data typically being shorter than $\sim$~1.8~ks) and the corresponding effect on the light-curve window function.

\subsubsection{SSS turn-on and turn-off times}

The turn-on and turn-off times of the SSS emission (and, hence, nuclear burning) can provide useful information about the WD parameters. With the high cadence monitoring regularly performed by \emph{Swift}, tying down these times more accurately becomes significantly easier.

For example, visibility of the supersoft emission requires that the nova ejecta are optically thin to X-rays. Therefore, the turn-on time of the SSS phase (together with the expansion velocity) tells us about the mass of the ejected shell.
\cite{schwarz11} reviewed a sample of \emph{Swift}-observed novae with supersoft emission, clearly demonstrating how the novae with faster velocities and an earlier SSS turn-on are consistent with lower ejected masses \citep{shore08}. Their results also show that recurrent novae tend to be located at the high-velocity/early-SSS/low-ejecta locus of the diagram, as expected for higher-mass WDs. 

Knowing the start/stop times of the SSS emission and, therefore, the duration of the nuclear burning interval, can also provide an estimate of the mass ejected \citep{shara10}.

\section{Open questions}
\label{open}

\cite{jpo15} presented a review of nova observations obtained by \emph{Swift} before 2015 April, including a list of open questions. Despite the four years which have passed since that work was published, to some extent the same areas for future work still remain. For completeness, we summarise and update these points below.

\begin{itemize}

\item
  This paper has concentrated mainly on temporal information obtained from \emph{Swift} observations. There are, of course, corresponding spectra obtained throughout the evolution of the novae. The problem which then arises is how to model these SSS data. While, at first glance, the spectra look decidedly blackbody-like, sometimes with superimposed absorption edges, from a physical standpoint this is not valid. \cite{kr96} first pointed out that parameterising the SSS spectra with blackbodies tends to underestimate the photospheric temperature, and overestimate the luminosity \citep[although this does not always seem to be the case;][]{jpo11b}. High-resolution grating data, from \emph{XMM-Newton} or \emph{Chandra}, reveal complex spectra \citep[e.g., ][]{ness07b} for which stellar atmosphere models are clearly required, and \cite{pageconf14} presented a preliminary investigation of \emph{Swift} spectra fitted with atmosphere grids. The resolution of CCD spectra such as those from \emph{Swift}-XRT is not typically high enough to constrain fully the parameters needed for such models. However, even when considering grating spectra, the atmosphere models available at the present time are not sufficiently advanced to parameterise the data accurately, and should be regarded as a work in progress.

\item  While QPOs have been detected in \emph{Swift} light-curves since 2006, our understanding of what causes them remains incomplete. Periodicities ranging from $\sim$~35--70~s have been seen in \emph{Swift}-XRT data during the SSS phase, in RS~Oph \citep{jpo06b, apb08b, jpo11a}, KT~Eri \citep{apb10}, V339~Del \citep{beard13} and V5668~Sgr \citep{page15c, gehrz18}. Possible explanations for these oscillations include pulsations or rotation of the WD. However, recent pulsation stability analysis suggests that those longer than 10--20~s for a high-mass system like RS~Oph would be quickly damped \citep{wolf18}. Considering rotation, periods under 100~s are not uncommon in cataclysmic variables \citep[e.g.,][]{ritter03}. However, assuming SSS emission originates from an extended nuclear-burning atmosphere, asymmetries would be required to produce modulations. In the case of magnetic WDs, hot spots could be caused by material funnelling into the poles, enhancing the nuclear burning in these regions; indeed, \cite{aydi18b} proposed that V407~Lup was a new IP nova system. Work to be included in a paper by Beardmore et al. finds that RS~Oph XRT data extracted during the intervals of the 35~s QPO show the spectrum to be harder at the time of the modulation maximum, which can be explained by variations in the oxygen column density. A similar change in column density was noted by \cite{ness15a} when exploring longer timescale variations found in \emph{Chandra} data.

\item  The link -- if any -- between the shock (non-SSS) emission in the XRT band and the \emph{Fermi}-LAT detections deserves further investigation. At the time of writing, 14 novae have been announced as having a GeV detection by the \emph{Fermi} Large Area Telescope (LAT); of these, nine are included in the \emph{Swift} sample presented here (Table~\ref{table:lat}), up until the end of 2017. The three more recent novae with LAT detections are V357 Mus \citep[also known as Nova Mus 2018 or PNV J11261220-6531086;][]{li18a}, V392~Per \citep[Nova Per 2018;][]{li18b} and V906~Car \citep[also known as ASASSN-18fv or Nova Car 2018;][]{jean18} -- which was also detected by AGILE \citep{piano18}. In addition, the earlier nova V1324 Sco (Nova Sco 2012) was also seen in the $\gamma$-rays by the LAT \citep{ack14}, but was never detected in the X-ray band by \emph{Swift} \citep{page12c, pagejpo13}; V5856 Sgr (ASASSN-16ma; Nova Sgr 2016 No. 4) had a LAT detection \citep{li16a, li16b, li17b}, but was only weakly seen in hard X-rays (no SSS emission) by \emph{Swift} (two observations were taken, 14.9 and 149.0 days after the nova discovery), so not included in this paper. \cite{franck18} also lists V1535~Sco (included in this \emph{Swift} sample) and V679~Car (Nova Car 2008; undetected by \emph{Swift}) as $\gamma$-ray emitting candidates, with significances of around 2$\sigma$. Of this sample of LAT novae, V745~Sco and V1534~Sco (and V1535~Sco with the tentative detection) are symbiotic (also recurrent, in the case of V745~Sco) novae, while the others are `standard' classical novae. Note that \emph{Fermi} was launched on 2008 June 11, meaning that the earliest 11 novae in this \emph{Swift} sample (Table~\ref{table:outburst}) were not observed by LAT around outburst.

  As more novae are detected at GeV energies, it appears possible that all novae are potential $\gamma$-ray sources. Given the relatively few GeV photons detected for any given nova, it is however likely that only nearby explosions would be detected, as discussed by \cite{morris17}. \cite{martdub13} consider the LAT detection of V407~Cyg, presenting a model where the $\gamma$-rays are caused by shock-accelerated electrons, while \cite{shore13} also conclude that internal shocks within the ejecta lead to the GeV emission in V959~Mon. The presence of shocks in nova systems is also revealed through radio data \citep[e.g.,][]{chom14b, weston16a}; \cite{metz14, metz15} propose that the reprocessing of early X-ray shocks by a dense, external shell may contribute significantly towards the optical/UV emission of novae. Early observations by \emph{Swift} and \emph{NuSTAR} \citep[Nuclear Spectroscopic Telescope Array;][]{harrison13} around the optical peak could provide additional information about this parameter space.

 \item Models of nova outbursts predict that there will be a short (0.5+ day) X-ray flash just after hydrogen ignition, but before the optical outburst. The duration of such a flash would be an indicator of the WD mass and accretion rate. However, catching a short flare in X-rays before the nova becomes optically bright requires either advance warning of an outburst (e.g., a recurrent nova with a well-known recurrence period), or all-sky X-ray surveys. \cite{morii16} placed limits on X-ray flashes for 40 novae using the Monitor of All-sky X-ray Image \citep[MAXI;][]{mats09}; however, the energy band of the Gas Slit Camera at 2--4~keV is likely too high to detect the expected SSS emission. The most rapidly recurrent nova known to date, M31N 2008-12a \citep[][and references therein]{henze18}, has a recurrence timescale of close to 1~yr, making it a suitable candidate for monitoring for precursive flashes. A high cadence (approximately every 6~hr) monitoring campaign was carried out by \emph{Swift} during the 8-day run-up to the eventual 2015 outburst \citep{kato16}. No X-ray emission was detected in this interval, with the conclusion by \cite{kato16} being that the expected flash probably occurred $\sim$~15.5 day pre-outburst, due to rapidly recurrent novae having lower maximum nuclear burning luminosities (despite their expected high WD masses). Future wide-field missions, such as \emph{Einstein Probe} \citep{yuan18}, will hopefully detect these expected X-ray flashes, which can then be followed up with \emph{Swift}.

\end{itemize}

\begin{table*}
  \begin{center}
\caption{Novae discussed in this paper which have a \emph{Fermi}-LAT detection.}
\begin{tabular}{ll}
  \hline
  Nova & LAT detection reference \\
  \hline
  V407 Cyg &  \cite{abdo10}\\
  V959 Mon &  \cite{cheung12}\\
  V339 Del & \cite{hays13}\\
  V1369 Cen  & \cite{cheung13a}\\
  V745 Sco & \cite{cheung14}\\
 V5668 Sgr &  \cite{cheung15,cheung16a}\\
 V407 Lup  & \cite{cheung16b}\\
 V5855 Sgr & \cite{li16}\\
  V549 Vel & \cite{li17a}\\
 
\hline
\end{tabular}
\label{table:lat}
\end{center}
\end{table*}

\section{Summary}

Since its launch in 2004, \emph{Swift} has observed in detail a large number of novae, leading to some interesting and unexpected results as highlighted here.
This paper presents novae from 2006--2017, bringing together the \emph{Swift} X-ray and UV light-curves of the well-monitored objects, almost all of which had detected supersoft emission. This dataset would lend itself to future statistical studies, where the X-ray properties could be compared between the different speed-classes of the novae, for example.

\emph{Swift} data are immediately public, and can be accessed online, at \\ http://www.swift.ac.uk/archive/ql.php, within a few hours of the observation taking place; the data are then moved to the archive at \\ http://www.swift.ac.uk/swift\_live/ around a week later. An online XRT product generator is provided by the UK Swift Science Data Centre (UKSSDC) at http://www.swift.ac.uk/user\_objects/ which can be used to generate spectra, light-curves and images \citep{evans07, evans09}.

\section{Acknowledgements}

These observations would not have been possible without the support of the \emph{Swift} PI (Neil Gehrels, and now Brad Cenko), together with the Mission and Flight Operations Teams. This work is presented on behalf of the \emph{Swift} Nova-CV group\footnote{http://www.swift.ac.uk/nova-cv/}, co-ordinated by J.P. Osborne, which is open to applications to join from all interested scientists. The \emph{Swift} project at the University of Leicester is funded by the UK Space Agency.



\end{document}